# Calculation of the Ion-Ion Recombination Rate Coefficient via a Hybrid Continuum- Molecular Dynamics Approach

Tomoya Tamadate[1,2], Hidenori Higashi[2,3], Takafumi Seto[2,3], Christopher J. Hogan Jr.[1*]

[1]Department of Mechanical Engineering, University of Minnesota, 111 Church St SE, Minneapolis, MN, 55455
[2]Faculty of Natural System, Graduate School of Natural Science and Technology, Kanazawa University, Kanazawa, Japan
[3]Faculty of Frontier Engineering, Institute of Science and Engineering, Kanazawa University, Kanazawa, Japan

[*]To whom correspondence should be addressed: E-mail: hogan108@umn.edu, T: 1-612-626-8312

**ABSTRACT**

Accurate calculation of the ion-ion recombination rate coefficient has been of long-standing interest, as it controls the ion concentration in gas phase systems and in aerosols. We describe the development of a hybrid continuum-molecular dynamics approach to determine the ion-ion recombination rate coefficient. The approach is based on the limiting sphere method classically used for transition regime collision phenomena in aerosols. When ions are sufficiently far from one another, ion-ion relative motion is described by diffusion equations while within a critical distance, molecular dynamics (MD) simulations are used to model ion-ion motion. MD simulations are parameterized using the AMBER force-field as well as by considering partial charges on atoms. Ion-neutral gas collisions are modeled in two mutually exclusive cubic domains composed of $10^3$ gas atoms each, which remain centered on the recombining ions throughout calculations. Example calculations are reported for $NH_4^+$ recombination with $NO_2^-$ in He, across a pressure range from 10 kPa to 10,000 kPa. Excellent agreement is found in comparison of calculations to literature values for the 100 kPa recombination rate coefficient ($1.0 \times 10^{-12}$ $m^3$ $s^{-1}$) in He. We also recover the experimentally observed increase in recombination rate coefficient with pressure at sub-atmospheric pressures, and the observed decrease in recombination rate coefficient in the high pressure continuum limit. We additionally find that non-dimensionalized forms of rate coefficients are consistent with recently developed equations for the dimensionless charged particle-ion collision rate coefficient based on Langevin dynamics simulations.

**Symbol Dictionary**

*Constants*

| | | | |
|---|---|---|---|
| $e$ | Electron charge | $1.60 \times 10^{-19}$ | [C] |
| $k_b$ | Boltzmann constant | $1.38 \times 10^{-23}$ | [J/K$^{-1}$] |
| $T$ | Temperature | 298.15 | [K] |
| $\varepsilon_0$ | Permittivity of vacuum | $8.85 \times 10^{-12}$ | [Fm$^{-1}$] |

*Variables*

| | | |
|---|---|---|
| $a_{ij}$ | Collision radius of ion $i$ and $j$ | [m] |
| $b$ | Collision parameter | [m] |
| $b_c$ | Critical collision parameter | [m] |
| $D_i$, $D_j$ | Diffusion coefficient of ion $i$ and $j$ | [m$^2$s$^{-1}$] |
| $E$ | Electric field | [Vm$^{-1}$] |
| $H$ | Dimensionless collision kernel | [-] |
| $I$ | Ion flow | [s$^{-1}$] |
| $I_\delta$ | Ion flow into limiting sphere surface | [s$^{-1}$] |
| $J$ | Ion flux density | [m$^{-2}$s$^{-1}$] |
| $Kn_D$ | Diffusive Knudsen number | [-] |
| $Kn_\delta$ | Diffusive Knudsen number for limiting sphere | [-] |
| $m_i$, $m_j$ | Mass of ion $i$ and $j$ | [kg] |
| $m_{ij}$ | Reduced masses of ion $i$ and $j$ $=m_i m_j/(m_i+m_j)$ | [kg] |
| $n_i$, $n_j$ | Ion $i$ and $j$ concentrations | [m$^{-3}$] |
| $n_\infty$ | Ion $i$ bulk concentrations | [m$^{-3}$] |
| $N_{col}$ | Number of collisions observed in the simulation | [-] |
| $N_{obs}$ | Number of orbits observed in the simulation | [-] |
| $N_{tot}$ | Total number of simulated trajectories | [-] |
| $P$ | Pressure | [Pa] |
| $p_\delta$ | Collision probability for ions entering the limiting sphere | [-] |
| $p_\chi$ | Collision probability for ions entering the trapping sphere | [-] |
| $p_F$ | Fuchs's collision probability | [-] |
| $r$ | Radial position (ion-ion distance) | [m] |
| $r_{km}$ | Radial distance for atoms $k$ and $m$ | [m] |
| $r_m$ | Minimum radial distance (ion-ion distance) | [m] |
| $t$ | Time | [s] |
| $v_0$ | Velocity at the limiting sphere surface | [m s$^{-1}$] |
| $v_e$ | Electrophoretic velocity at the limiting sphere surface | [m s$^{-1}$] |
| $v_{th}$ | Thermal velocity at the limiting sphere surface | [m s$^{-1}$] |
| $z_k, z_m$ | Partial charge of specie $k$ and $m$ | [-] |
| $\beta_{ij}$ | The collision rate coefficient for type $j$ ions with type $i$ ions which have arrived at $a_{ij}$ | [m$^3$ s$^{-1}$] |
| $\beta_\delta$ | The collision rate coefficient for type $j$ ions with type $i$ ions which have arrived at $\delta$ | [m$^3$ s$^{-1}$] |
| $\chi$ | Trapping sphere radius | [m] |
| $\delta$ | Limiting sphere radius | [m] |
| $\varepsilon_{km}$ | Lenard-Jones parameter between atoms $k$ and $m$ | [J] |
| $\Phi_{inter}$ | Intermolecular potential | [J] |
| $\Phi_{LJ}$ | Lenard-Jones potential | [J] |
| $\Phi_E$ | Electric potential | [J] |
| $\gamma$ | Enhancement factor | [-] |
| $\eta_c, \eta_f$ | Dimensionless continuum regime and free molecular enhancement factors | [-] |
| $\lambda_{ij}$ | Ion-ion mean free path | [m] |
| $\theta_0$ | Incident angle at limiting sphere surface | [-] |
| $\theta_c$ | Critical angle for collision | [-] |
| $\theta_{c,\chi}$ | Critical angle to enter trapping sphere | [-] |
| $\sigma_{km}$ | Lenard-Jones parameter for atoms $k$ and $m$ | [m] |
| $\Psi_E$ | Magnitude of the Coulombic energy to the thermal energy ratio at the limiting sphere surface | [-] |

## I. Introduction

Ion-ion recombination refers to instances where ions of opposing polarity collide with one another, and via charge transfer reactions the products are neutral vapor phase species.[1-4] Recombination is crucially important in many gas phase and aerosol systems; the rate of recombination relative to the rate of ion formation determines ion concentrations in the ambient [5,6], in high temperature combustion systems [7], and in intentionally irradiated systems utilized in modulating the charge on aerosol particles (i.e. in aerosol chargers[8,9]). Ion concentrations, in turn, can affect gas phase reaction and ambient particle formation rates[10], and via the charging of particles, particle-particle interactions[11-13] and particle measurements.[14,15] Gas phase and aerosol systems are thus highly sensitive to ions and ion-ion recombination.

The recombination rate itself is the product of the number concentrations of the colliding ions, $n_i$ and $n_j$ for ion species $i$ and $j$, respectively, and the recombination rate coefficient, $\beta_{ij}$ (which is typically denoted with the symbol $\alpha$, but which we denote as $\beta_{ij}$ for reasons noted subsequently). Recombination is typically considered to be a collision-controlled process, hence $\beta_{ij}$ is determined by the motion of two ions about one another until collision. While on the surface this appears to be a relatively simple process, calculation and measurement of the recombination rate has been a topic of scientific interest for more than a century,[2,4,16-18] and improved recombination rate coefficient calculations still remain of interest. Along these lines the purpose of this work is to develop a new method of calculating the ion-ion recombination rate coefficient, accounting precisely for the influences of ion structures, ion-ion potential interactions, and ion-neutral background gas collisions on the recombination process, at variable temperatures and pressures.

To better justify development of a new calculation approach, we first summarize prior ion-ion recombination rate coefficient measurements and calculations, noting why specifically improved calculation methods are important. Under tropospherically relevant conditions, measurements repeatedly suggest a recombination rate coefficient of order 1.0-3.0 x $10^{-12}$ $m^3$ $s^{-1}$.[1, 4, 17, 19-21] In general, the recombination rate coefficient increases with increasing pressure at sub-atmospheric levels, and increases with decreasing temperature.[5-7, 22-24] Further observations are dependencies on the ion chemical composition and background gas relative humidity (which also affects ion composition[5, 25]), as well as a decrease in the rate with increasing pressure at elevated, super-atmospheric pressures.[4, 20, 26] Calculation approaches that can accurately predict how the recombination rate coefficient changes above or below atmospheric pressure, at variable temperature, gas composition, and ion composition are still lacking. The earliest calculation approaches relevant to atmospheric ions were performed by Thomson.[2] Discussed in detail by Loeb & Marshall,[4] Thomson modeled recombination such that even at low pressure, ion recombination is influenced by three-body trapping, i.e. as two oppositely charged ions migrate about one another, there is a non-negligible probability that one of them collides with a neutral gas molecule, and this ion-neutral collision alters the ion trajectory such that an ion-ion collision subsequently occurs when otherwise it would not. Three-body trapping can be contrasted with the two-body collision approach, where ion-neutral gas collisions are negligible, and which is widely utilized in modeling high energy electron collisions with positive ions and positively charged particles and probes in low pressure plasmas.[16, 27] It can also be contrasted with the continuum (Langevin) approach, where ion motion is resisted by multiple neutral gas collisions,[18] as would be the case in a high pressure gas. Calculation approaches to bridge the combined two-body collision, three-body trapping low pressure regime and the continuum, high pressure regime were later developed by Natanson,[28] Brueckner,[1] and Bates & Flannery,[29] all treating ions as structureless point entities. Subsequent calculation efforts were devoted to modeling this process using Monte-Carlo simulations[30, 31] and to scrutinizing implementation of combined three-body trapping- two-body collision models near atmospheric conditions.[24] Overall, while agreement between theoretical calculations, Monte-Carlo simulations, and measurements can be achieved (often with some tuning of input values), no single calculation approach has proven applicable over a wide range of ion chemical compositions, temperatures, pressures, and gas compositions (see these references[5, 24, 25, 32] for examples of disagreement between theory and measurements).

The issues with prior calculation approaches arise largely because it is difficult first to analytically model ion-ion relative motion across a wide pressure range, and second to model ion-neutral gas collisions without precisely accounting for ion-neutral interactions and instead treating ions as structureless point charges. To accurately capture ion-neutral gas collisional influences, molecular dynamics (MD) simulations, or at least gas molecule scattering calculations[33, 34] need to be carried out. Such calculations need to consider realistic ion and gas molecule structures (all-atom models), accurate ion-neutral potential interactions, the appropriate gas molecule kinetic energy (velocity) distribution for the temperature of interest, and a procedure to accurately describe gas molecule impingement and rebound during collision with an ion. While the solution would seem to be to simply apply detailed MD simulations to model ion-ion recombination, even with modern computational power, it is generally not feasible to carry out complete MD simulations modeling ion-ion motion in a domain large enough to circumscribe the region where ions Coulombically interact with one another. With this issue in mind, to implement MD simulations to precisely examine ion trajectories on close approach to one another and the possibility of multiple ion-neutral gas molecule collisions, here we develop and implement a hybrid continuum-molecular dynamics approach, wherein motion is modeled via continuum equations when ions are sufficiently far from one another, and via MD simulations as ions approach one another. Such approaches have been developed previously to investigate non-continuum fluid flow[35]; here the approach is simplified in that the continuum portion of the model can be treated analytically and the continuum model and the molecular dynamics model can be solved decoupled from one another.

In combining spatially separated regions where ion migration is modeled via continuum and non-continuum approaches respectively, our approach builds upon the limiting sphere models of particle ionization (particle-ion collisions) of Fuchs[36] and more directly, Filippov[37], who originally described a universal continuum-non-continuum limiting sphere model that is readily implementable with MD simulations. In utilizing an approach applicable to ion-ion collisions as well as particle-ion and particle-particle collisions, we also attempt to unify theories describing binary collision-limited reactions via a single calculation framework (hence utilizing $\beta_{ij}$ as the recombination coefficient, as it is typically used to denote the collision rate coefficient for all possible colliding partners in aerosols). In the section that follows, we describe the equations utilized in the hybrid model. A test case is presented of $NH_4^+$ recombination with $NO_2^-$ in a variable pressure Helium environment ($10^1$ kPa – $10^4$ kPa) at 300 K. This test case is largely chosen for computational simplicity for initial model development (a monoatomic noble gas). Results are compared to the near atmospheric pressure recombination rate measurements of Lee & Johnsen[25]; this comparison shows excellent agreement near atmospheric pressure, where measurements were performed.

## II. Theory & Numerical Methods

We first present the limiting sphere approach of Fillipov,[37] which, although not the original work presenting the limiting sphere approach, provides the most general form for this model, and forms the basis for the approach utilized in this study (section II.A). Prior to discussing MD calculations, we also provide a review of Fuchs's[36] assumptions in recombination rate coefficient and particle-ion collision rate coefficient calculation via the limiting sphere approach (section II.B), as well as a presentation of Hoppel & Frick's[38] modifications to Fuchs's approach (section II.C.). We elect to review these two, alternative approaches because we believe full presentation of these theories and their limitations better motivates the development of a new calculation method. We additionally compare MD calculation results to predictions from these theories. Nonetheless, readers not concerned with the calculation details of prior theories may proceed to second II.D following section II.A., with minimal loss in scope. Following discussion of MD simulations in section II.D., we then discuss non-dimensionalization of the recombination rate coefficient (section II.E.).

### A. Limiting Sphere Theory

We consider a domain where ion $j$ is positioned at the center, and the relative motion of type $i$ ions is monitored about this central ion (Figure 1). The limiting sphere radius, $\delta$, is defined such that at distances beyond it, relative motion can be described fully by continuum diffusion equations. Therefore, the spatial concentration evolution of ions of type $i$ beyond $\delta$ obeys the equation:

$$\frac{\partial n_i(r)}{\partial t} + \frac{\partial J}{\partial r} = 0 \tag{1}$$

where:

$$J = -(D_i + D_j)\frac{\partial n_i(r)}{\partial r} + \frac{e}{k_b T}(D_i + D_j)E(r)n_i(r) \tag{2}$$

$n_i$ is the concentration of type $i$ ions, $t$ is time, $r$ is radial position (ion-ion distance), $e$ is the unit electron charge, $D_i$ is the diffusion coefficient of ion $i$, $k_b$ is Boltzmann's constant, $T$ is temperature, and $E$ is the electric field formed by the ions. In equation (2) both ions are assumed singly charged. In the present study, we also assume a simple Coulomb form for the electric field outside the limiting sphere radius, $E(r) = \frac{-e}{4\pi\varepsilon_0 r^2}$ as short range potential interaction terms are negligible at long distances ($\varepsilon_0$ is the permittivity of free space). Combining equations (1) and (2) and assuming that the concentration profile rapidly reaches a steady-state yields:

$$(D_i + D_j)\frac{\partial n_i(r)}{\partial r} + (D_i + D_j)\frac{e^2}{4\pi\varepsilon_0 k_b T r^2}n_i(r) = A \tag{3}$$

where $A$ is a constant. The total flow of type $i$ ions to any radial coordinate equal to or larger than the limiting sphere radius is then defined as:

$$I = 4\pi r^2 A = 4\pi r^2 (D_i + D_j)\frac{\partial n_i(r)}{\partial r} + (D_i + D_j)\frac{e^2}{\varepsilon_0 k_b T}n_i(r) \tag{4}$$

Equation (4) is a first order differential equation, whose solution for the boundary condition $n_i \to n_\infty$ as $r \to \infty$ was determined by Fuchs[36] to be:

$$I = 4\pi(D_i + D_j)\frac{n_\infty - n_i(r)exp\left(\frac{-e^2}{4\pi\varepsilon_0 k_b T r}\right)}{\int_r^\infty \frac{1}{x}exp\left(\frac{-e^2}{4\pi\varepsilon_0 k_b T x}\right)dx} = \frac{(D_i+D_j)e^2}{\varepsilon_0 k T}\frac{n_\infty - n_i(r)exp\left(\frac{-e^2}{4\pi\varepsilon_0 k_b T r}\right)}{1-exp\left(\frac{-e^2}{4\pi\varepsilon_0 k_b T r}\right)} \tag{5}$$

Equation (5) would apply to the point of collision (i.e. $r = a_{ij}$, where $a_{ij}$ is the point of ion-ion contact) if continuum transport equations were valid for all ion-ion separation distances. Therefore, equation (5) can be applied up to a point $r = \delta$ and then equated with the product of the type $i$ ion concentration at $\delta$ and the collision rate coefficient for type $j$ ions with type $i$ ions which have arrived at $\delta$, $\beta_\delta$:

$$I_\delta = \frac{(D_i+D_j)e^2}{\varepsilon_0 kT} \frac{n_\infty - n_i(\delta)\exp\left(\frac{-e^2}{4\pi\varepsilon_0 k_b T\delta}\right)}{1-\exp\left(\frac{-e^2}{4\pi\varepsilon_0 k_b T\delta}\right)} = n_i(\delta)\beta_\delta \tag{6}$$

Filippov[37] showed that this limiting sphere collision rate coefficient can be calculated via the equation:

$$\beta_\delta = \frac{4\delta^2 p_\delta}{(2-p_\delta)}\left(\frac{2\pi k_b T}{m_{ij}}\right)^{1/2} \tag{7}$$

where $m_{ij}$ is the reduced mass of ions and $p_\delta$ is the probability that a type $i$ ion entering the limiting sphere will collide with ion $j$. The recombination coefficient, $\beta_{ij}$ can then be defined by rearrangement of equation (6). Defining $\beta_{ij} = \frac{I_\delta}{n_\infty}$, and noting that $n_i(\delta) = \frac{I_\delta}{\beta_\delta} = \frac{\beta_{ij}n_\infty}{\beta_\delta}$ yields:

$$\beta_{ij} = \frac{(D_i+D_j)e^2}{\varepsilon_0 kT\left(1-\exp\left(\frac{-e^2}{4\pi\varepsilon_0 k_b T\delta}\right)\right)}\left(1 + \frac{(D_i+D_j)e^2}{\beta_\delta \varepsilon_0 k_b T\left(\exp\left(\frac{e^2}{4\pi\varepsilon_0 k_b T\delta}\right)-1\right)}\right)^{-1} \tag{8a}$$

$$\beta_{ij} = \frac{4\pi(D_i+D_j)\delta\Psi_\delta}{(1-\exp(-\Psi_\delta))}\left(1 + \left(\frac{\pi}{2}\right)^{1/2}\frac{(2-p_\delta)}{p_\delta}Kn_\delta \frac{\Psi_\delta}{(\exp(\Psi_\delta)-1)}\right)^{-1} \tag{8b}$$

$$\Psi_\delta = \frac{e^2}{4\pi\varepsilon_0 k_b T\delta} \tag{8c}$$

$$Kn_\delta = \left(\frac{m_{ij}}{k_b T}\right)^{1/2}\frac{(D_i+D_j)}{\delta} \tag{8d}$$

In equation (8c), $\Psi_\delta$ is the magnitude of the Coulombic energy to the thermal energy ratio at the limiting sphere surface, and in equation (8d), $Kn_\delta$ is a diffusive Knudsen number[39] for the limiting sphere. Though equations (8b-d) are uniquely represented here (i.e. we do not believe these equations have been presented in this manner previously), their derivation follows the approach of Fillipov[37] and as noted by Fillipov, equation (8a) is similar to that derived by Fuchs [36] in developing the limiting sphere model. Fillipov[37] additionally notes that equation (8) can be implemented for any selected value of δ, provided that beyond $\delta$ the continuum transport approximation is valid; in fact, the continuum transport approximation can remain valid within a portion of δ, provided that a method to accurately determine $p_\delta$ for properly chosen values of δ is implemented. For this reason, equations (8a-d) are applied with the MD simulation approach discussed in section II.D.

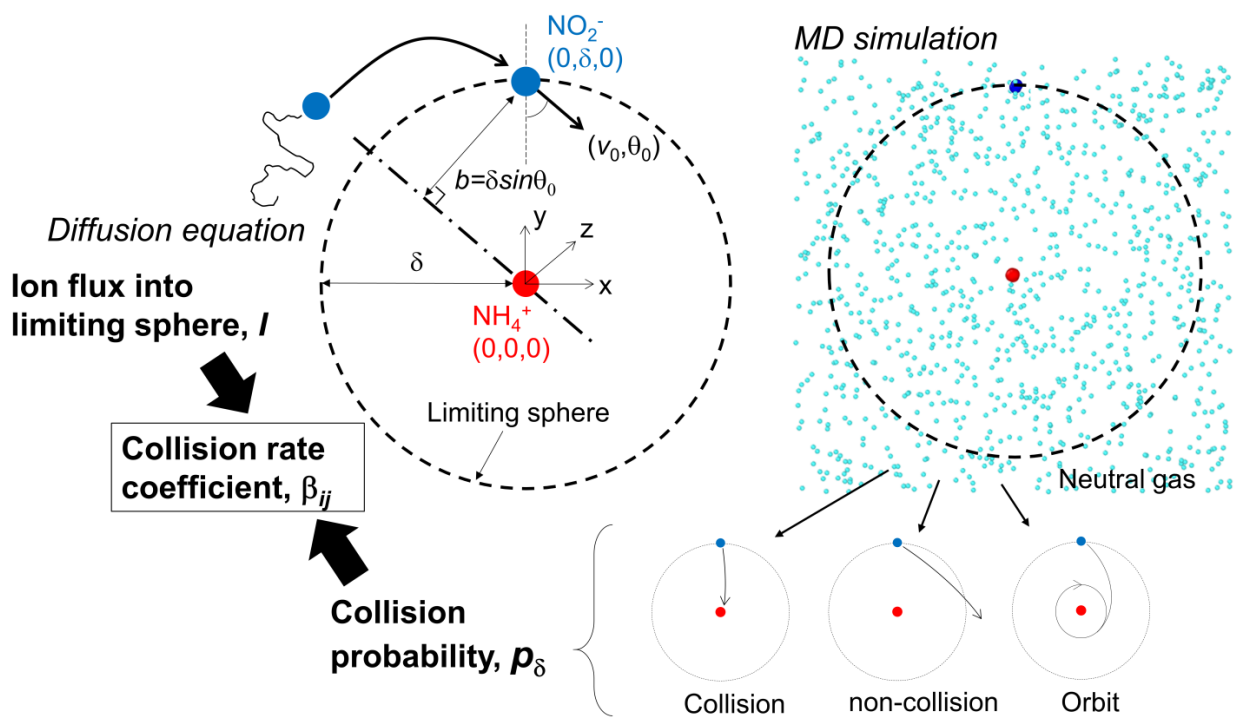

**Figure 1.** A depiction of ion motion to the limiting sphere surface (upper left). Determination of the probability of collision for an ion initiated on the limiting sphere surface (entering ion) with an ion initiated at the center (center ion) of the sphere yields the ion-ion recombination rate via equation (8). Molecular Dynamics simulations (depicted in the upper right) can be used in lieu of traditional theories to compute the probability of collision accounting for ion-neutral gas encounters. Ion-ion motion yields either collision, non-collision, or orbit (lower line) and any theory defining the probability of collision must explicitly state how each event is defined and accounted for collision probability analysis.

## B.. Fuchs's Assumptions in Probability Calculation

In developing and utilizing the limiting sphere approach to determine particle-ion collision rate coefficients (with extension to ion-ion recombination coefficients), Fuchs[36] utilized equation (8a) with very specific assumptions. First, following the discussion of Wright,[40] he selected the limiting sphere radius as:

$$\delta = \frac{a_{ij}^3}{\lambda_{ij}^2}\left[\frac{1}{5}\left(1+\frac{\lambda_{ij}}{a_{ij}}\right)^5 - \frac{1}{3}\left(1+\frac{\lambda_{ij}^2}{a_{ij}^2}\right)\left(1+\frac{\lambda_{ij}}{a_{ij}}\right)^3 + \frac{2}{15}\left(1+\frac{\lambda_{ij}^2}{a_{ij}^2}\right)^{5/2}\right] \quad (9a)$$

where $a_{ij}$ is the collision radius (the point at which the two ions or ion and particle collide), and $\lambda_{ij}$, commonly referred to as the ion-ion mean free path, is given as:

$$\lambda_{ij} = (D_i + D_j)\left(\frac{\pi m_{ij}}{8 k_b T}\right)^{1/2} \quad (9b)$$

Second, Fuchs assumed that ion relative motion inside the limiting sphere can be described completely by free molecular (collisionless) kinetics, hence $\beta_\delta$ can be expressed as:

$$\beta_\delta = \pi \delta^2 \left(\frac{8 k_b T}{\pi m_{ij}}\right)^{1/2} p_F \quad (10)$$

Equation (10) is a free molecular collision rate coefficient (projected area and mean thermal speed product) multiplied by the probability ($p_F$) that an ion impinging upon limiting sphere does in fact collide with the central ion.

To calculate $p_F$, Fuchs[36] utilized simplified trajectory calculations for point ions (with a charged particle at the center), wherein ions were initiated on the limiting sphere surface with a single speed $v_0$. Conservation of energy and conservation of angular momentum during ion migration yield:

$$\frac{1}{2}m_{ij}v_0^2 - \frac{e^2}{4\pi\varepsilon_0\delta} = \frac{m_{ij}}{2}\left[\left(\frac{dr}{dt}\right)^2 + r^2\left(\frac{d\theta}{dt}\right)^2\right] - \frac{e^2}{4\pi\varepsilon_0 r} \quad (11)$$

$$bm_{ij}v_0 = m_{ij}r^2\left(\frac{d\theta}{dt}\right). \quad (12)$$

where $r$ and $\theta$ are the radial and angular coordinates of the incoming ion (with the central ion remaining fixed in the examined frame of reference), $b = \delta\sin\theta_0$ is the initial impact parameter (Figure 1), $\theta_0$ is the initial entering angle, and $\frac{dr}{dt}$ and $\frac{d\theta}{dt}$ are the radial and angular velocities inside the limiting sphere, respectively. Combining these two equations yields the ion trajectory upon entering the limiting sphere as:

$$\left(\frac{dr}{d\theta}\right)^2 = \left(\frac{r^2}{b}\right)^2\left\{1 - \left(\frac{b}{r}\right)^2 + \frac{e^2}{2\pi m_{ij}v_0^2\varepsilon_0}\left(\frac{1}{r} - \frac{1}{\delta}\right)\right\} \quad (13)$$

The condition $\frac{dr}{d\theta} = 0$ yields a relationship between the minimum radial distance ($r_m$) that the two species under examination approach one another and their initial impact parameter:

$$b = r_m\sqrt{1 + \frac{e^2}{2\pi m_{ij}v_0^2\varepsilon_0}\left(\frac{1}{r_m} - \frac{1}{\delta}\right)} \quad (14a)$$

Equation (14a) can be then used to define a critical impact parameter ($b_c$) for collision by equating $r_m$ with the ion-ion collision radius $a_{ij}$:

$$b_c = a_{ij}\sqrt{1 + \frac{e^2}{2\pi m_{ij}v_0^2\varepsilon_0}\left(\frac{1}{a_{ij}} - \frac{1}{\delta}\right)} \quad (14b)$$

For all impact parameters less than $b_c$, corresponding to $r_m < a_{ij}$, collision would occur (given Fuchs's assumptions). Additionally the assumption that ions enter the limiting sphere uniformly distributed in initial $\theta_0$ yields:

$$p_F = \left(\frac{b_c}{\delta}\right)^2 \tag{15}$$

From equations (10), (14b) and (15), a final form for $\beta_\delta$ is obtained as:

$$\beta_\delta = \pi a_{ij}^2 \left(\frac{8k_bT}{\pi m_{ij}}\right)^{1/2} \gamma(a_{ij}), \tag{16a}$$

$$\gamma(a_{ij}) = 1 + \frac{e^2}{4\pi\varepsilon_0 k_bT}\left(\frac{1}{a_{ij}} - \frac{1}{\delta}\right). \tag{16b}$$

In $\gamma(a_{ij})$, $kT$ is used in place of $\frac{1}{2}m_{ij}v_0^2$. This substitution ensures that the equation (8a) recombination coefficient converges to the ballistic (free molecular) limit expression correctly. We note that Fuchs[36] utilized $\frac{3}{2}kT$ instead of $kT$, leading to incorrect ballistic limit (low pressure) two-body calculations.

Equations (16a-b), with equation (8a), yield predictions for the ion-ion recombination coefficient with little difficulty in computation. However, the development of Fuchs's approach requires several assumptions rendering it invalid in ion-ion recombination coefficient prediction, and its predictions can differ by an order of magnitude from literature reported ion-ion recombination rates. First, while there is no issue with the definition of a critical radius $\delta$ beyond which a continuum transport approximation is valid, it is not correct to assume that within $\delta$, ion-neutral gas molecule collisions negligibly affect their trajectories, i.e. it is not correct to completely neglect three-body trapping. This assumption was a major concern of Hoppel & Frick[38] in the development of a limiting sphere theory including three-body trapping, as well as the concern of others in expanding upon their work.[41-43] Second, as noted by Gopalakrishnan & Hogan[44] but hitherto unaddressed in limiting sphere based approaches, on the limiting sphere boundary, incoming ions are not uniformly distributed in $\theta_0$ and their initial speed distribution function at the limiting sphere boundary will deviate significantly from a thermal equilibrium distribution. The change in velocity and angle distributions is attributed to the influence of potential interactions on ion motion prior to entering the limiting sphere; ions are attracted to one another and hence accelerated to much higher velocities in the direction of their center-to-center vector. To demonstrate this latter point, we note that in the Cartesian coordinate system depicted in Figure 1, wherein the "y" direction is aligned with the ion center-to-center vector, at δ, the initial relative velocity vector between ions is given as $\vec{v_0} = (v_x, v_y, v_z) = (v_{th}, v_{th} + v_e, v_{th})$, where $v_{th}$ is sampled from the Maxwell-Boltzmann 1D distribution function:

$$f(v_{th}) = \sqrt{\frac{m_{ij}}{2\pi k_b T}} exp\left(-\frac{m_{ij} v_{th}^2}{2 k_b T}\right) \qquad (17a)$$

and $v_e$ is the average electrophoretic velocity at δ:

$$v_e = \frac{(D_i + D_j)}{k_b T} \frac{e^2}{4\pi\varepsilon_0 \delta} \qquad (17b)$$

Example plots of the joint probability density function (pdf, denoted as $\frac{\partial^2 n^*}{\partial v_0 \partial \theta_0}$) for the initial ion speed ($v_0 = \|\vec{v_0}\| = \sqrt{v_x^2 + v_y^2 + v_z^2}$) and the initial angle $\theta_0 = \cos^{-1}\left(\frac{v_y}{v_0}\right)$ for the collision of $NO_2^-$ and $NH_4^+$ in He background gas at 300 K and variable pressures are shown in Figure 2. These pdfs were determined via randomly sampling $10^7$ velocities per plot using equation (17a), rejecting all trajectories which would not enter the limiting sphere. The properties of $NO_2^-$ and $NH_4^+$ under these conditions are given in Table 1; these properties are used in all reported calculations. δ in Figure 2 is calculated via equation (9a) using a collision radius of $a_{ij} = 3.35$Å. For example calculations, the diffusion coefficients were calculated using the equations of Fuller et al;[45] while this technically applies to the neutral vapor molecule, this approximation does not have a substantial influence on the presented results as the test gas (helium) is negligibly polarizable at 300 K. The test conditions yield a relative mean thermal speed between the two ions of 700 m s$^{-1}$. At the lowest pressures, where the limiting sphere radius is largest, potential interactions do not strongly influence the initial speed and angle distribution. However, as pressure increases, the decreasing value of δ leads to large increases in the mean speed, shifting the joint pdf to the right, and leading to preferential motion in the y direction. For comparison, lines denoting $\theta_c$ are displayed, where $\theta_c$ is defined from equation (14b) via the relationship $b_c = \delta \sin\theta_c$.

**Table 1.** A summary of the ion properties utilized in calculations.

|  | Nitrogen dioxide ion, $NO_2^-$ | Ammonium ion, $NH_4^+$ |
|---|---|---|
| Nominal Molecular mass [Da] | 46 | 18 |
| Diffusion Coefficient*Pressure [m$^2$/s·Pa] | 7.02 | 7.42 |
| Electrical mobility*Pressure [m$^2$/Vs·Pa] | 273 | 289 |
| Approximated Radius [Å] | 1.775 | 1.575 |

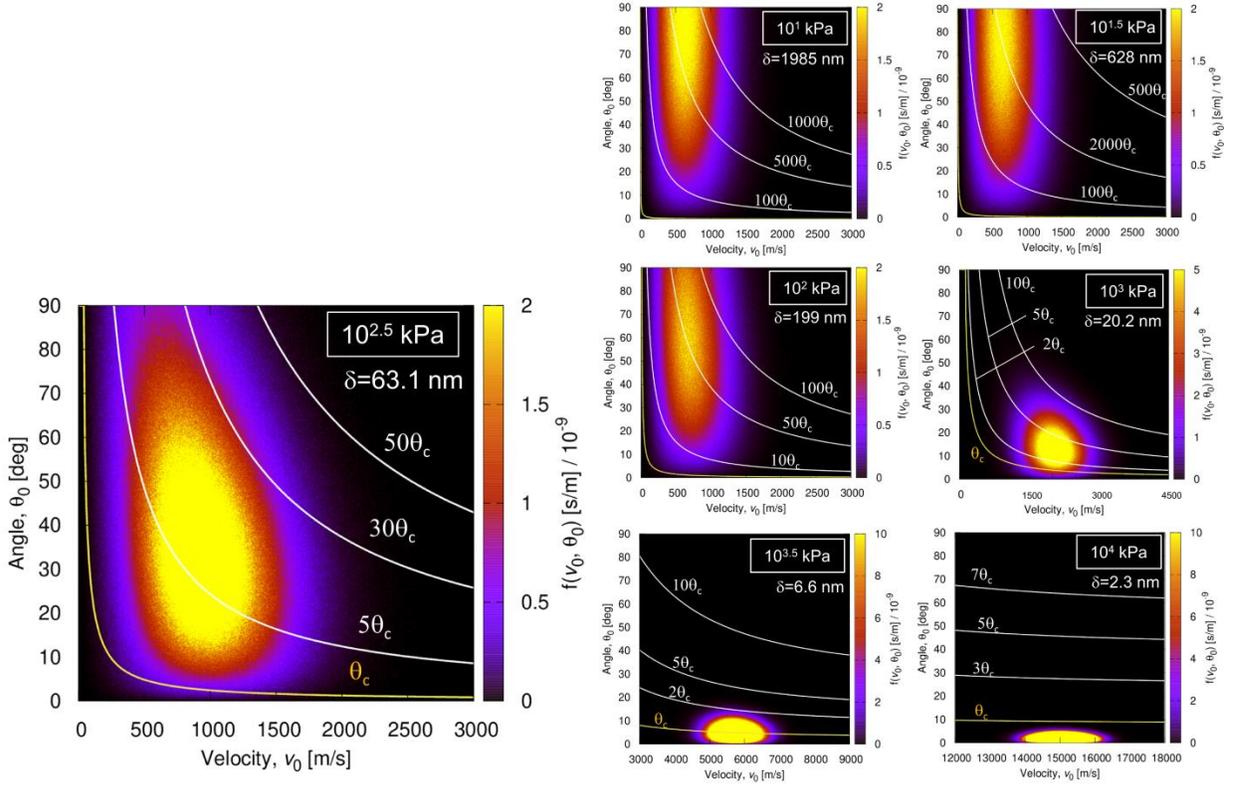

**Figure 2.** Heat maps showing the joint probability distribution functions (pdfs, $\frac{\partial^2 n^*}{\partial v_0 \partial \theta_0}$) for the initial angle ($\theta_0$, as per Figure 1) and initial relative speed for ions entering a Fuchs limiting sphere, considering $NO_2^-$ (entering ion) recombining with $NH_4^+$ (center ion). In contrast to these heap maps, traditional limiting sphere calculation approaches in aerosol particle charging assume either the entering ion speed is at the mean thermal speed (relative) or Maxwell-Boltzmann distributed in speed, with a uniform angle distribution. The true joint pdf varies with pressure (upper right of each heat map) because of how the limiting sphere radius varies with pressure. $\theta_c$ denotes the critical angle in the Fuchs[36] model for collision as a function of velocity; for $\theta_0 < \theta_c$, collision occurs.

## C. Hoppel & Frick's Probability Modification

To account for three-body trapping in limiting sphere calculations, Hoppel & Frick[38] invoked the concept of a trapping sphere of radius $\chi$; if the center-to-center distance of the entities under examination becomes less than $\chi$, collision would occur with probability $p_\chi$.

Redefining the limiting sphere radius as: $\delta = \lambda_{ij} + \chi$, they proposed that $\beta_\delta$ (equation 10) can be expressed as:

$$\beta_\delta = \pi\chi^2 \left(\frac{8k_bT}{\pi m_{ij}}\right)^{1/2} \gamma(\chi) p_\chi \tag{18}$$

where $\gamma(\chi)$ is calculated via equation (16b), replacing $a_{ij}$ with $\chi$. Equation (18) and the equations presented subsequently would apply in instances where $\chi > a_{ij}$; otherwise, Hoppel & Frick[38] suggest use of equation (16a) using their modified $\delta$ definition. Implementation of their modification hence requires calculation of $p_\chi$ and $\chi$. For a given $\chi$, it can be shown that:

$$p_\chi = 1 - \frac{\lambda_{ij}^2}{2\chi^2}\left[1 - exp\left(\frac{-2\chi cos\theta_{c,\chi}}{\lambda_{ij}}\right)\left(1 + \frac{2\chi}{\lambda_{ij}}cos\theta_{c,\chi}\right)\right] \tag{19a}$$

$$\theta_{c,\chi} = sin^{-1}\left(\frac{b_{c,\chi}}{\chi}\right) \tag{19b}$$

$$b_{c,\chi} = a_{ij}\sqrt{1 + \frac{e^2}{32k_bT\varepsilon_0}\left(\frac{1}{a_{ij}} - \frac{1}{\chi}\right)} \tag{19c}$$

Unfortunately, Hoppel & Frick[38] do not provide a means to determine $\chi$. As they were specifically concerned with particle-ion collision rate coefficients, they proposed $\chi$ can be determined using the ion-ion recombination rate coefficient itself at a given pressure and temperature. With the recombination rate ($\beta_{input}$) known, they proposed that the theoretical expression of Natanson[28] can be employed to determine an ion-ion trapping sphere radius, from which a separate particle-ion trapping sphere radius can be determined. In the present study, the ion-ion recombination rate is the parameter of interest, hence $\chi$ would be the ion-ion trapping sphere radius itself. According to Natanson[28], the input ion-ion recombination rate is linked to this trapping distance via the expression:

$$\beta_{input} = \frac{\pi\chi^2\left(\frac{8k_bT}{\pi m_{ij}}\right)^{1/2} f(g)\left[1+\frac{e^2\lambda_{ij}}{4\pi\varepsilon_0 k_b T\chi(\chi+\lambda_{ij})}\right]exp\left[\frac{e^2}{4\pi\varepsilon_0 k_b T(\chi+\lambda_{ij})}\right]}{1+\frac{\pi\varepsilon_0 k_b T\chi^2 f(g)}{e^2\lambda_{ij}}\left[1+\frac{e^2\lambda_{ij}}{4\pi\varepsilon_0 k_b T\chi(\chi+\lambda_{ij})}\right]\left\{exp\left[\frac{e^2}{4\pi\varepsilon_0 k_b T(\chi+\lambda_{ij})}\right]-1\right\}} \qquad (20a)$$

where function $f(g)$ is:

$$f(g) = 2\omega - \omega^2 \qquad (20b)$$

$$\omega = 1 + 2\left|\frac{exp(-g)}{g^2} + \frac{exp(-g)}{g} - \frac{1}{g^2}\right| \qquad (20c)$$

$$g = \frac{2\chi}{\lambda_{ij}} \qquad (20d)$$

While it first glance implementation of equations (20a-d) to infer $\chi$ and use of equation (18) to yield $\beta_\delta$ may seem tractable, the end result is a circular argument; a recombination rate coefficient is required as an input to determine the recombination rate coefficient. Furthermore, Hoppel & Frick[38] do not discuss the consequence of changing the system pressure on the trapping distance. Pressure changes most certainly affect the ion-ion recombination rate[20, 26], the value of $\lambda_{ij}$, and likely change the most appropriate estimate of $\chi$. However, if $\chi$ must be recalculated for changes in temperature and pressure, the ion-ion recombination rate would need to be recalculated in the first place to again determine the ion-ion trapping distance. For this reason we believe this approach cannot truly be employed used to predict the ion-ion recombination rate *a priori*, and is best thought of as a fitting procedure wherein a single value of $\chi$ is determined for a single $\beta_{input}$ (at a single temperature and pressure), and subsequently this value is used to predict the recombination rate coefficient or particle-ion collision rate coefficient for variable pressure and temperature conditions. We also remark that neither should a complete theory predicting the ion-ion recombination rate (or particle-ion recombination rate) require the trapping distance as an input nor is the trapping distance a physically rigorous parameter. The

continuum – molecular dynamics hybrid approach obviates the need to discuss a hypothetical three-body trapping distance.

*D. Molecular Dynamics based Probability Calculation*

We now return to equations (8b-d), which simply require a means of determining $p_\delta$ for implementation. To utilize MD simulations to compute $p_\delta$ we position ion $i$ (the incoming ion) at a Cartesian coordinate location $(0,\delta,0)$ and position ion $j$ (the central ion) at $(0,0,0)$. The equations of motion are then solved for both ions. The initial center-of-mass velocity vector of ion $i$ is sampled from the joint probability density functions in Figure 2 (i.e. using equations 17a & 17b), while ion $j$ is modeled at rest initially. The individual atoms in each ion have initial velocities which are the sum of the initial center-of-mass velocity and an additional thermal term, determined via separate NVT equilibration simulations at 300 K. We carried out calculations both including and omitting the influences of equation (17b), i.e. with and without the influence of electrostatic forces on the incoming ion initial velocity. We first discuss MD simulations carried out in the absence of neutral gas. Potential interactions ($\phi_{\text{inter}}$) between atoms within different ions are modeled via the Lennard-Jones 6-12 potential with the long range electrostatic potential:

$$\phi_{\text{inter}} = \phi_{\text{LJ}} + \phi_{\text{E}} = 4\epsilon_{km}\left[\left(\frac{\sigma_{km}}{r_{km}}\right)^{12} - \left(\frac{\sigma_{km}}{r_{km}}\right)^{6}\right] + \frac{z_k z_m e}{4\pi\varepsilon_0 r_{km}} \qquad (21)$$

where $\epsilon_{km}$ and $\sigma_{km}$ are the Lenard-Jones potential parameters between atom $k$ and atom $m$, $z_k$ is the partial charge of atom $k$, and $r_{km}$ is the atom-atom center-to-center distance. Lennard-Jones parameters for $NH_4^+$ and $NO_2^-$ were selected based on the AMBER force-field[46], and the partial charges were determined via NIH (National Institute of Health) database values. Intramolecular bond and angle potentials are additionally included, and were also calculated here based on the AMBER force-field. The AMBER force-field was simply selected as it is well-studied force-field; however, we remark that the influence of force-field choice on collision-controlled gas phase reactions is less studied than for molecular dynamics in solution, and this choice may need to be more carefully scrutinized in future continuum-molecular dynamics implementations. As depicted in Figure 1, calculations in the absence of background gas proceeded until one of three outcomes occurred: (1) the ion-ion separation distance was less than the collision radius $a_{ij}$ (collision); (2) the ion-ion separation distance exceeded δ (non-collision); or (3) the ions were observed to enter stable orbit about one another. To compute $p_\delta$ we made sure to monitor at least 100 case (1) collision events, and $p_\delta$ was determined both treating orbits as non-collision events ($p_\delta = N_{col}/N_{tot}$, where $N_{col} = 100$ is the number of collision events and $N_{tot}$ is the total number of trajectory cases examined) and treating orbits as collision events ($p_\delta = (N_{col} + N_{orb})/N_{tot}$, where $N_{orb}$ is the number of orbiting trajectories observed).

Calculations neglecting neutral gas need only consider a small number of atoms (< 10), hence they proceed efficiently irrespective of the size of δ selected. However, equation (9a) calculated limiting sphere radii of $10^2$ nm and higher can lead to instances where the number of gas molecules which would occupy the domain bounded by the limiting sphere exceeds $10^4$ (and approaches $10^7$). As $p_\delta$ is anticipated to decrease to well below unity (by orders of magnitude) with decreasing pressure, a greater number of trajectories must also be modeled for accurate calculation at low pressure; combined this makes MD simulations considering all gas molecules in the limiting sphere domain computationally intractable across a wide pressure range. To circumvent this issue while accounting for neutral gas, we instead opt to surround each ion with a cubic domain containing $10^3$ neutral He gas atoms and whose dimensions is determined by the background gas temperature and pressure (hence simulating only 2000 additional atoms). All He atoms are initially positioned at random locations within their assigned cubic domain, with random velocity directional vectors and random speeds sampled from the Maxwell-Boltzmann speed distribution at the input temperature. Within each domain, He atom motion is monitored via the velocity-Verlet algorithm, with a number of simplifications. First, He-He interactions are completely ignored. This assumption has no bearing on ion trajectories except at the highest pressures examined, where the gas mean free path approaches the ion size, but as $p_\delta \to 1$ in the high pressure limit (the continuum limit), even at high pressure this approximation has no bearing on MD calculations (though the diffusion coefficients of ions would need to be modified in equation (8)). Second, He atoms interact with all atoms within the ion in their assigned box via Lenard-Jones potential interactions (i.e. equation 17a omitting the Coulombic term). Helium-$NH_4^+$ and Helium-$NO_2^-$ potential parameters were estimated from the Lorentz-Berthelot combination rule[47], with $\epsilon_{He-He}$=0.0203 kcal/mol $\sigma_{He-He}$ =2.556 Å$^2$. The cubic domains are moved throughout trajectory calculations such that they remain centered on their respective ions. Importantly, He atoms do not interact with atoms in the ion not in their assigned box; therefore as ions approach one another and their domains intersect, the effective gas density around each ion remains constant. When a He atom leaves its assigned domain, periodic boundary conditions are invoked; while this leads to a modest increase in the total kinetic energy in each simulation (as the ions accelerate one another) we find that including $10^3$ atoms leads to this increase negligibly influencing results. Third, because gas atoms only influence ion motion via collision (close range interaction), gas atoms which are a distance of 5.0 nm or more from their assigned ion's center-of-mass are moved with a variable time step shown in the supporting information, while those within this radial distance are moved with the ion's time step of 1 fs. Lennard-Jones potential interactions were not calculated for atoms more than 1.5 nm from one another.

Combined, the three aforementioned model assumptions yield tractable $p_\delta$ calculations for ion-ion recombination at pressures as low as 10 kPa (and likely lower, but at greater computational expense). They additionally enable detailed accounting of the effect of ion-neutral gas collisions, obviating the need to assume either specular or diffuse gas atom scattering upon collision, i.e. all degrees of freedom in ions and gas atoms are modeled[48, 49]. This is an important feature in ion-ion recombination calculations, as the model required to accurately depict gas atom or molecule scattering from an ion without modeling ion internal degrees of freedom (frozen ion models) has been found strongly dependent on ion chemical composition and size.[50, 51] An example trajectory calculation including He gas atoms is depicted in Figure 3, with variable orientation views provided. Videos depicting trajectories resulting in collision, non-collision, and orbit are also including as supporting information files (where gas atoms are not shown). The line trace in Figure 3 depicts the trajectory of the incoming ion relative to the central ion (repositioned to remain fixed at the domain center). Qualitatively, the trajectory displays features which cannot result in the absence of background gas; the reversals in trajectory direction are the result of ion-neutral collisions. Simulations with gas atoms are also carried out until $10^2$ collision events are observed; in the presence of neutral gas atoms, stable orbits are not observed.

Calculations do require input values of δ and $a_{ij}$, yet the calculation result needs to be independent of these choices; with neutral gas considered, δ needs to be sufficiently large, while $a_{ij}$ needs to be sufficiently small such that ions cannot escape one another at this distance. We probe the influence of both of these parameters, and report on these sensitivity analyses in the *Results and Discussion* section (section III). The reported calculations were carried out using hardware both at the Research Center for Computational Science (Okazaki, Japan) and the Minnesota Supercomputing Institute (MSI) of the University of Minnesota. A custom written C-code was employed for all calculations using trajectory calculation algorithms based upon the Large-scale Atomic/Molecular Massively Parallel Simulator (LAMMPS) source code,[52] which is open-source and optimized for efficient trajectory calculations.

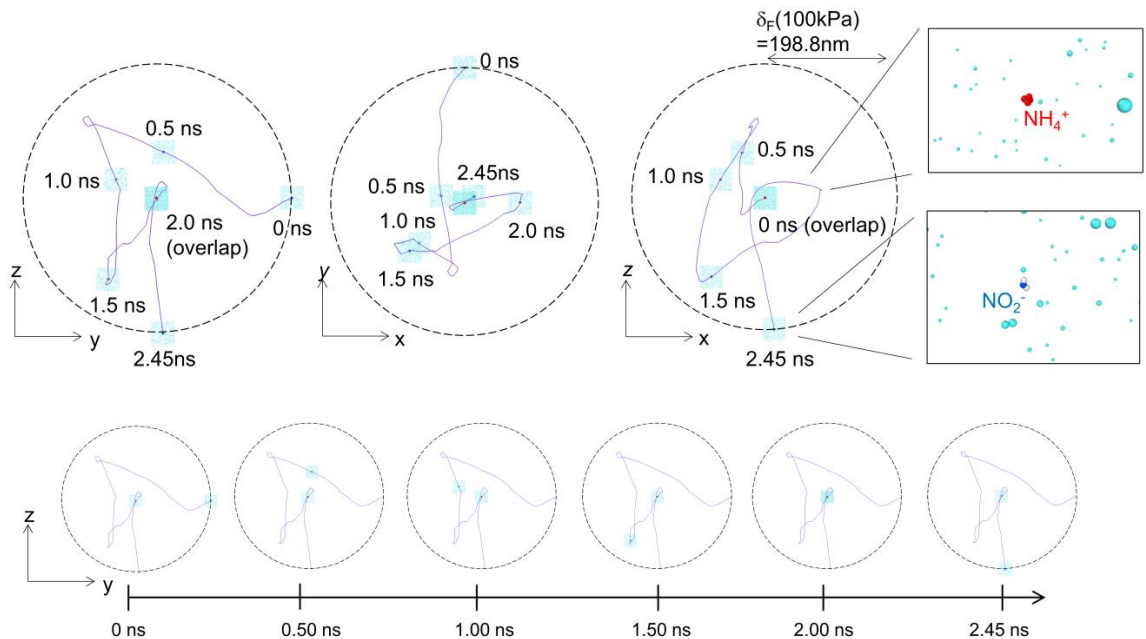

**Figure 3.** Depiction of relative ion trajectories during a single molecular dynamics simulation (100 kPa). Each ion (center ion: $NH_4^+$, entering ion: $NO_2^-$) is surrounded by $10^3$ He atoms which do not interact with one another but fully interact with the atoms in the ion they are assigned to. The displayed instance does not result in collision.

*E. Non-Dimensionalization*

As a final point in recombination rate coefficient method development, we are also interested in examining collapsed, dimensionless forms of the recombination rate coefficient. A series of recent studies have shown that the dimensionless collision rate coefficient ($H$) in particle-particle,[39, 53, 54] particle-vapor molecule,[55] and particle-ion[44, 56, 57] collisions can be expressed, almost universally, as a function the diffusive Knudsen number ($Kn_D$), with these two parameters defined as:

$$Kn_D = \frac{\sqrt{m_{ij}k_bT}\eta_c}{f_{ij}a_{ij}\eta_f} \tag{22a}$$

$$H = \frac{\beta_{ij}m_{ij}\eta_c}{f_{ij}a_{ij}^3\eta_f^2} \tag{22b}$$

where $f_{ij}$, defined as $(D_i+D_j)/k_bT$ for two ions, is the effective reduced friction factor. $\eta_c$ and $\eta_f$ are the dimensionless continuum regime and free molecular enhancement factors, respectively, which take different functional forms based upon the potential interactions between the two colliding species under consideration. Neglecting shorter range interactions, for attractive Coulomb interactions between single charged ions alone $\eta_c$ and $\eta_f$ are respectively expressed as:

$$\eta_c = \frac{\Psi_E}{1-exp(-\Psi_E)} \tag{23a}$$

$$\eta_f = 1 + \Psi_E \tag{23b}$$

$$\Psi_E = \frac{e^2}{4\pi\varepsilon_0 k_bT a_{ij}} \tag{23c}$$

Gopalakrishnan & Hogan[39] developed a regression equation to Langevin dynamics simulations for hard sphere collisions, showing that:

$$H_{HS}(Kn_D) = \frac{4\pi Kn_D^2 + 25.836 Kn_D^3 + \sqrt{8\pi}\,11.211 Kn_D^4}{1 + 3.502 Kn_D + 7.211 Kn_D^2 + 11.211 Kn_D^3} \tag{24}$$

where the subscript "HS" denotes the hard-sphere curves. Interestingly, via appropriate calculation of $\eta_c$ and $\eta_f$, this curve has been found exendable to short range, attractive potentials (van der Waals and image potentials) and to repulsive Coulomb interactions.[13, 57] However, equation (24) yields incorrect results for attractive Coulomb interactions outside the $Kn_D \to 0$ (continuum) and $Kn_D \to \infty$ (free molecular) limits.[44] Recently, Chahl & Gapalakrishnan[56] developed improved regressions for attractive Coulomb potentials over a wide $\Psi_E$ and $Kn_D$ range (0<$\Psi_E$<60, $Kn_D$<2000):

$$H(Kn_D, \Psi_E) = e^{\mu(Kn_D, \Psi_E)} H_{HS}(Kn_D) \tag{25a}$$

$$\mu(Kn_D, \Psi_E) = \frac{C}{A}\left(1 + k\frac{\ln(Kn_D) - B}{A}\right)^{-\frac{1}{k}-1} \exp\left[-\left(1 + k\frac{\ln(Kn_D) - B}{A}\right)^{-\frac{1}{k}}\right] \tag{25b}$$

$$A = 2.5 \tag{25c}$$

$$B = 4.528 \exp(-1.088\Psi_E) + 0.7091 \ln(1 + 1.537\Psi_E) \tag{25d}$$

$$C = 11.36\Psi_E^{0.272} - 10.33 \tag{25e}$$

$$k = -0.003533\Psi_E + 0.05971 \tag{25f}$$

We compare the dimensionless expression provided by equations (25a-f) to continuum-molecular dynamics determined recombination rate coefficients, non-dimensionalizing calculation results using equations (22a-b).

## III. Results & Discussion

### A. Limiting Sphere Radius Selection

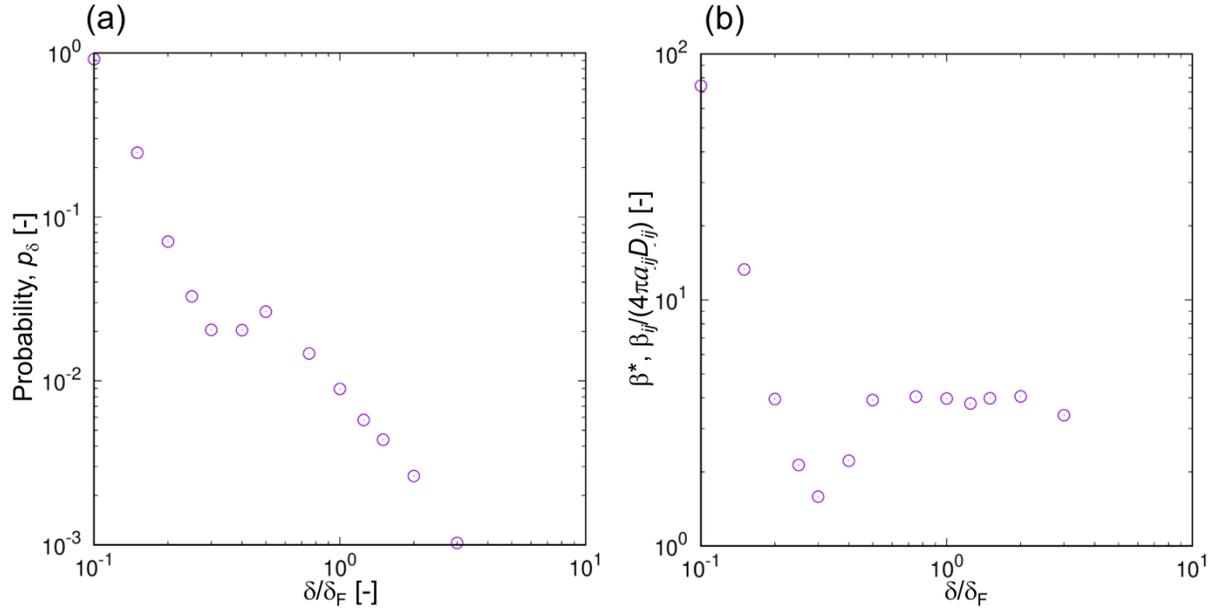

**Figure 4.** Resulting probability of collision **(a)** and recombination rate coefficient **(b)** for variable values of the limiting sphere radius, δ, normalized by the Fuchs[36] limiting sphere radius from equation (9a). The recombination rate coefficient is normalized by the hard-sphere continuum value.

In utilizing equation (9a), Fuchs[36] was attempting to identify a center-to-center distance beyond which continuum equations were approximately valid and within which free molecular trajectory calculations could be applied. The latter restriction does not apply in Fillipov's[37] approach and does not apply in continuum-molecular dynamics calculations; nonetheless, δ still needs to be sufficiently large for proper calculation. Near atmospheric pressure (100 kPa), we varied the choice of δ to examine its influence on calculations considering neutral gas and with the initial incoming ion velocity sampled as $\vec{v_0} = (v_{th}, v_{th} + v_e, v_{th})$. As a function of the ratio δ/δ$_F$, where δ$_F$ denotes the equation (9a) prediction (with $a_{ij} = 3.35$Å), Figure 4a displays the resulting $p_\delta$, while Figure 4b displays $\beta_{ij}^*$, the recombination rate coefficient calculated with equation (8b), normalized by $4\pi a_{ij}(D_i + D_j)$. With increasing $\delta/\delta_F$ beyond 0.5, we find that $p_\delta$ decreases in manner leading to a near constant value of $\beta_{ij}^*$, suggesting that values of $\delta$ larger than $\frac{1}{2}\delta_F$ are sufficient for calculations at atmospheric pressure. For limiting sphere radii selected smaller than $\frac{1}{2}\delta_F$ the assumption of purely continuum transport beyond the limiting sphere radius appears to break down, leading to large changes in $\beta_{ij}^*$. For the remainder of the calculations reported, based upon this sensitivity analysis at atmospheric pressure, we employ $\delta$ from equation (9a) with $a_{ij} = 3.35$Å.

B. *The Recombination Rate Coefficient without Neutral Gas*

Table S1 of the supporting information summarizes all continuum-molecular dynamics calculations performed. Prior to discussing calculation results with ion-neutral gas atom collisions, we first examine calculations in the absence of neutral gas, as in instances where the incoming ion velocity is sampled omitting electrostatic influences, results should agree with Fuchs's[36] theory. Figure 5(a) and Figure 5(c) display plots of $p_\delta$ for initial incoming ion velocities accounting for both thermal and electrostatic effects, and only accounting for thermal effects, respectively. Figures 5(b) and 5(d) show the resulting $\beta_{ij}$ calculations for $p_\delta$ results provided in Figures 5(a) and 5(c), respectively. The probability plots contain dashed lines for the values in the continuum limit ($p_\delta \to 1$) and free molecular limit. The latter is given by the expression:

$$p_F = \left(\frac{a_{ij}}{\delta}\right)^2 (1 + \Psi_E) \qquad (26)$$

We display results both discounting and including orbiting trajectories as collisions. When discounting orbiting trajectories, $p_\delta$ approaches the minimum of the continuum and free molecular limiting curves under all conditions, irrespective of whether the incoming ion velocity is sampled including or excluding electrostatic potential influences. This suggests that the initial velocity has little influence on non-orbiting ion-ion relative trajectories, presumably because the electrostatic potential terms in the equations of motion rapidly increase ion velocities to the appropriate values in the absence of neutral gas. Interestingly, Fuchs[36] calculations are in excellent agreement with our calculations when excluding orbiting trajectories, suggesting that the assumptions made in the original limiting sphere theory development are consistent with omitting the possibility of orbit.

Including orbiting trajectories as collisions drastically changes calculation results outside the continuum limit such that the deviations from Fuchs[36] predictions are substantial. Using calculation values at 100 kPa discounting and including orbiting trajectories to determine $\chi$ (smaller values correspond to discounting orbiting trajectories), the Hoppel & Frick[38] approach can be used to fit calculation results with reasonable agreement near the selected pressure used in fitting, for the orbiting case. However, because the Hoppel & Frick[38] model does not converge to the two-body free molecular limit as pressure decreases, it cannot capture the behavior of the orbit-excluded models, and it is a poorer fit outside the continuum limit when the electrostatic velocity is included as an initial condition. As stable orbits cannot be established in the presence of neutral gas (the ions will either eventually be pushed towards each other or pushed away from one another), the disparity in results including and discounting orbiting trajectories highlights the need to carry out full MD simulations within the limiting sphere.

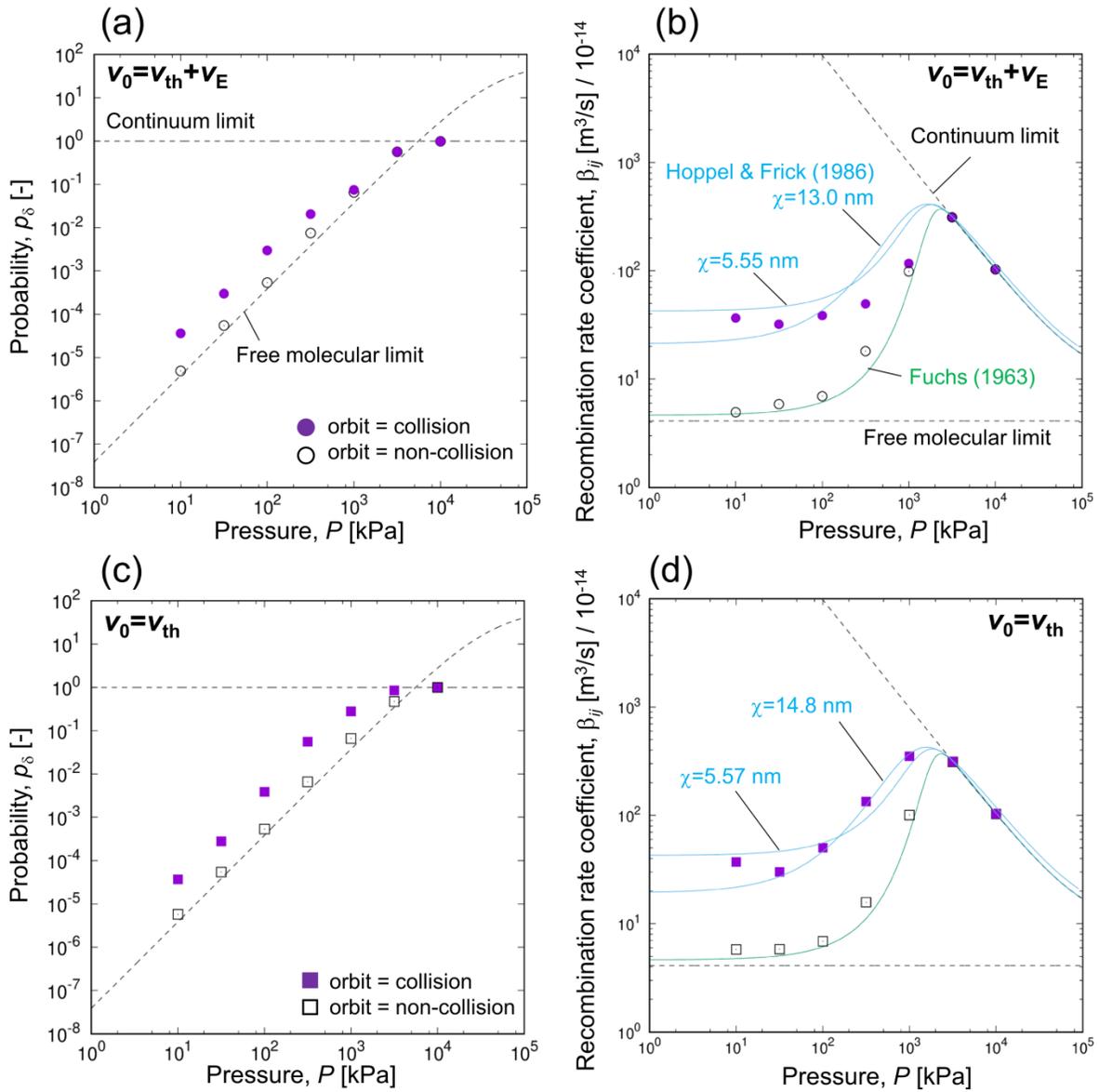

**Figure 5.** The probability of collision, **(a)** & **(c)**, and recombination rate coefficient, **(b)** & **(d)**, resulting from molecular dynamics calculations in the absence of neutral gas atoms. Dashed lines denote the continuum and free molecular limits in all plots. **(a)** & **(b)** correspond to ions initiated with initial velocities and angles accounting for both thermal and electrostatic effects, while **(c)** & **(d)** correspond to ions initiated with thermal effects only. Closed symbols denote orbiting cases considered as collisions, while open symbols denote results considering orbit ions as non-collision events. In **(b)** & **(d)** predictions of Fuchs[36] (equations 8a, & 16a-b, green lines), and of Hoppel & Frick,[38] (equations 8a, 16a-b, and 18, blue lines) are shown. The Hoppel & Frick[38] predictions are based upon the recombination rate coefficient at 100 kPa from simulations as an input, leading to the $\chi$ values shown in each plot; smaller values correspond to calculations discounting orbits in collisions.

## C. The Recombination Rate with Ion-Neutral Collisions

We compute $p_\delta$ using MD simulations by randomly sampling initial conditions for ions and gas molecules; we found that this is the most computationally efficient method to determine the collision probability, and fixing the number of collision events to be monitored at $10^2$ sets the counting statistic uncertainty in $p_\delta$ at 10% of $p_\delta$ itself. However, a more rigorous approach would be to compute $p_\delta(v_0, \theta_0)$, i.e. the collision probability for specific initial incoming ion speeds and angles. The value of $p_\delta$ would then be computed as:

$$p_\delta = \int_0^{\pi/2} \int_0^\infty p_\delta(v_0, \theta_0) \frac{\partial^2 n^*}{\partial v_0 \partial \theta_0} dv_0 d\theta_0 \tag{27}$$

While we do not employ equation (27), examination of $p_\delta(v_0, \theta_0)$ and the product $p_\delta(v_0, \theta_0) \frac{\partial^2 n^*}{\partial v_0 \partial \theta_0}$ calculated with and without neutral gas is of interest; this highlights how ion-neutral collisions influence ion trajectories. Figures 6 (a), (b), and (c) display plots of $p_\delta(v_0, \theta_0)$ calculated without neutral gas excluding orbit, without neutral gas including orbit, and with neutral gas, respectively, at 100 kPa. Such plots are similar to the collision probability curves of Yang et al.[58] and Halonen et al.[59] in examining impact parameters and initial speeds of molecules and clusters leading to condensation in the free molecular regime. Guidelines denote $\theta_c$, the critical angle for collision in Fuchs[36] theory (below this curve, collision occurs, above it, ions do not collide) and multiples of $\theta_c$. Excluding orbit yields $p_\delta(v_0, \theta_0)$ in line with Fuchs[36] definition of $\theta_c$, while including orbit shifts the $p_\delta(v_0, \theta_0)$ heat map slightly on a log-log scale. However, the inclusion of neutral gas molecules completely changes the $p_\delta(v_0, \theta_0)$ surface (we note the streaks in Figure 6c appear to be the result of spline fitting to a finite number of tested $v_0, \theta_0$ locations). Including neutral gas, collisions are neither certain for any initial speed or angle, nor are they prohibited at any initial speed-angle combination except large angle, large velocity initial trajectories (which almost immediately leave the limiting sphere after entry). This highlights the need to explicitly include gas atoms in detailed modeling of ion trajectories; complete exclusion of gas atoms and approximation of ion-neutral collision influences via a three-body trapping distance both cannot recover the $p_\delta(v_0, \theta_0)$ surfaces obtained from precise examination of ion-neutral collisions.

Figures 6 (d), (e), and (f) display heat maps of the product $p_\delta(v_0, \theta_0) \frac{\partial^2 n^*}{\partial v_0 \partial \theta_0}$ for the conditions corresponding to 6(a), 6(b), and 6(c), respectively. Heat maps highlight the most probable collision type for the prescribed calculation conditions. Through comparison of Figure 6(f) to 6(d) and 6(e), it is evident that inclusion of neutral gas shifts the initial angles leading to most collisions to larger values than are predicted to be possible using Fuchs[36] theory or anticipated based on the exclusion of neutral gas. Stated differentially, at atmospheric pressure most ion-ion collisions occur with ions initially in grazing collision trajectories which would not collide in the absence of gas.

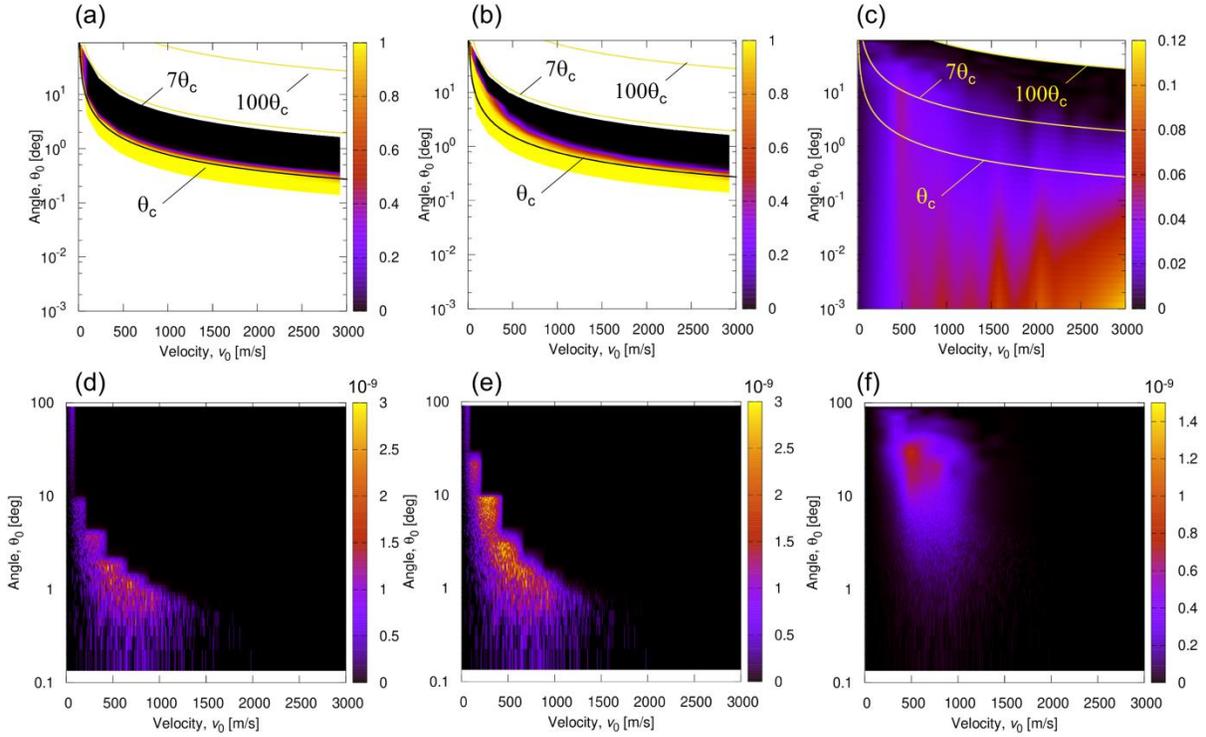

**Figure 6.** The probability of collision as a function of $\theta_0$ and $v_0$ at 100 kPa, for simulations without neutral gas atoms and ignoring orbiting events (**a**), without neutral gas atoms but counting orbiting as collision (**b**), and including ion-neutral collisions (**c**). For the three examined cases, the resulting products of the probability of collision and joint probability density function for the initial angle and velocity are shown in (**d**), (**e**), and (**f**), respectively.

Analogous to Figure 5, in Figure 7 we plot the probabilities of collision including neutral gas (Figure 7a) and the resulting recombination rate coefficient (in Figure 7b). Results in the absence of neutral gas are included for comparison, along with continuum and free molecular limiting lines. Gas inclusion increases the overall probability of collision, leading to recombination coefficients higher by an order of magnitude than the orbit excluded instances outside the continuum regime, and higher by a factor of ~2 than the orbit included instances outside the continuum regime. Including neutral gas near atmospheric pressure, we arrive at a recombination rate coefficient of 1.1 x $10^{-12}$ $m^3$ $s^{-1}$ including electrostatic effects in the initial velocity. Lee & Johnsen[25] report a recombination rate coefficient near 1.0 x $10^{-12}$ $m^3$ $s^{-1}$ in He near atmospheric pressure, which is in outstanding agreement with calculations. Near a pressure of 30 kPa they report a recombination rate coefficient near 7 x $10^{-13}$ $m^3$ $s^{-1}$; this is again in excellent agreement with our calculations (6.7 x $10^{-13}$ $m^3$ $s^{-1}$) as pressure is reduced below atmospheric levels. An increase in recombination rate with increasing pressure at low pressure, and then its decrease with increasing pressure at high pressure has been documented in some of the earliest recombination rate experiments,[26, 60] and it is this unique behavior that served as driver for many of the initial theoretical investigations of ion-ion recombination. Our calculations capture this phenomenon; calculations outside the continuum limit reveal a monotonically increasing recombination rate coefficient with increasing pressure and at the continuum limit, a decreasing recombination rate coefficient with increasing pressure. In total, we find that the explicit inclusion of neutral gas in calculations has a substantial influence on the recombination rate coefficient calculation, and that without any fit parameters (i.e. using input for the test ions and neutral gas inferred from their tabulated properties, without adjustment), we can achieve excellent agreement between continuum-molecular dynamics hybrid calculations and experimental measurements.

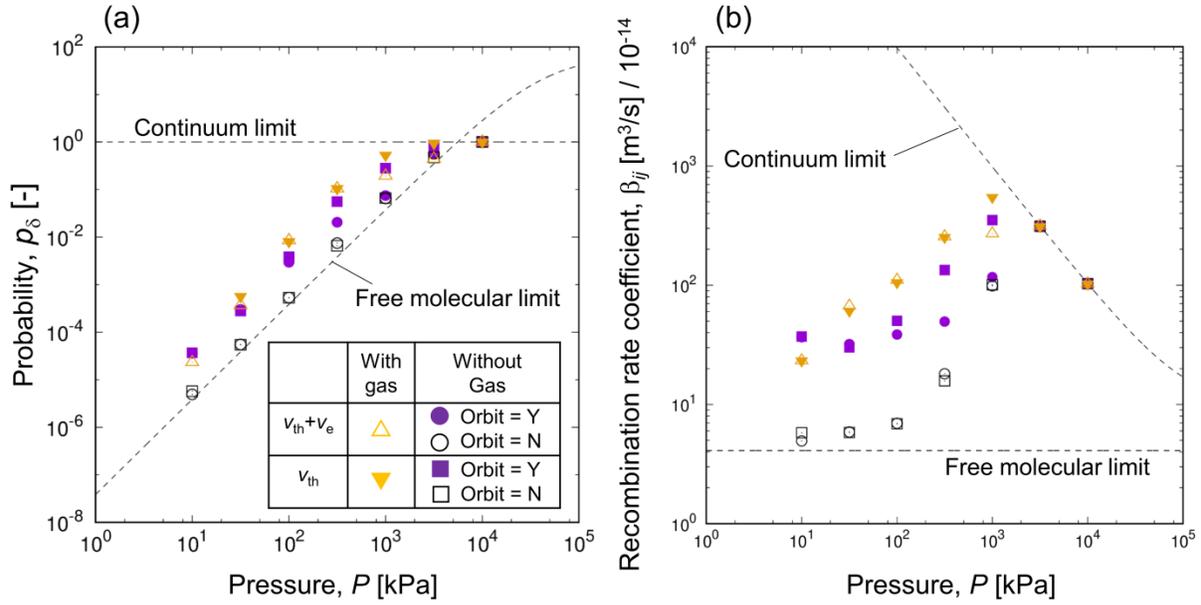

**Figure 7.** The probability of collision **(a)** and recombination rate coefficient **(b)**, both considering neutral gas (triangles) and neglecting neutral gas (circles, squares).

## D. Non-Dimensional Curves

A final pending question is the extent to which calculations agree with regression based dimensionless curves resulting from Langevin simulations. However, to non-dimensionalize results appropriately, the choice of the collision radius $a_{ij}$ needs to be carefully scrutinized, as neither ion is strictly spherical, and the collision radius, i.e. the center-to-center distance within which charge neutralization is certain, is not easily defined. For 100 kPa calculations including neutral gas and electrostatic potential effects on the entering ion velocity, in Figure 8 we plot the recombination rate coefficient as a function of the input collision radius. Alterations to the collision radius do not require new calculations; the same set of trajectories can be re-analyzed with a different assumed $a_{ij}$ value. The true collision radius will be the largest distance below which the collision radius assumed has no bearing on the inferred recombination rate coefficient. In Figure 8 the input value of 3.35Å is found to be far below the true collision radius of 12.8 Å. We thus elect to use 12.8 Å as the collision radius in non-dimensionlization.

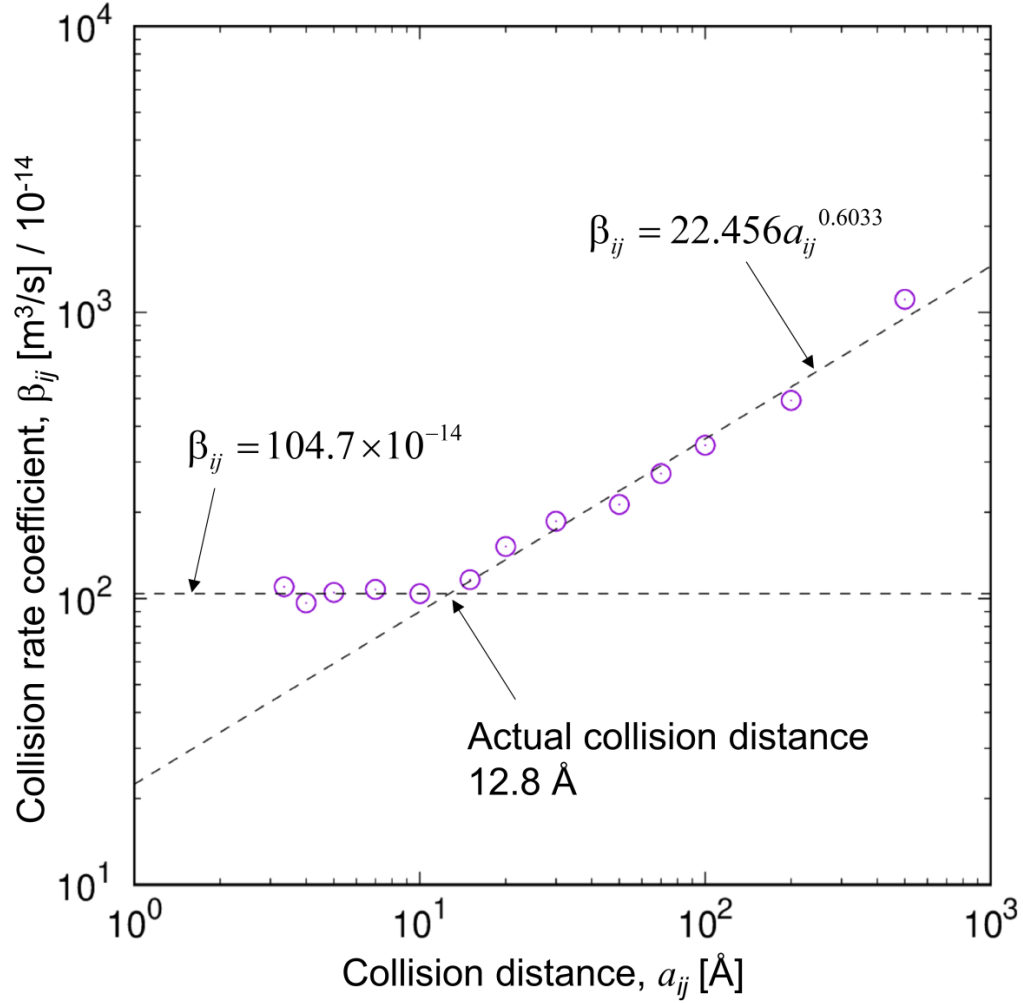

**Figure 8.** The recombination rate coefficient as a function of the input collision radius (center-of-mass to center-of-mass distance considered to be collision) at 100 kPa.

Figure 9 displays a plot of the dimensionless rate coefficient $H$ from equation (22b) as a function of $Kn_D$ from equation (22a) from calculations including neutral gas and using the inferred collision radius of 12.8 Å in non-dimensionalization. Also plotted are the continuum ($H = 4\pi Kn_D^2$) and free molecular ($H = \sqrt{8\pi}Kn_D$) limit expressions, non-dimensionalized Fuchs (1963) predictions, the Chahl & Gopalakrishnan[56] regression equation, and a non-dimensionalized Hoppel & Frick[38] fit where $\chi$ = 21.6 nm is based upon the recombination rate coefficient at 100 kPa. In large part, both the Chahl & Gopalakrishnan[56] regression equation and the Hoppel & Frick[38] fit capture our calculations well at $Kn_D$ < 100. Deviations begin to manifest at $Kn_D$ > 100 and are significant at $Kn_D$ > 1000. This high $Kn_D$ limit is near the upper bound of the Chahl & Gopalakrishnan[56] regression equation, and as discussed previously, the Hoppel & Frick[38] model does not converge appropriately to the two body collision limit, which will eventually be reached at low enough pressure, i.e. sufficiently high diffusive Knudsen number. While dimensionlessly, our results apply only to a specific value of $\Psi_E = 43.4$, results do suggest that with improvements, the dimensionless collision rate coefficient curves developed for particle collisions can be extended to ion-ion recombination, which vastly simplifies recombination rate estimation. At the same time, when higher precision calculations are needed or the influences of ion structure are of interest, continuum-molecular dynamics hybrid calculations appear to be a tractable method to study ion-ion collision phenomena in gases and aerosols.

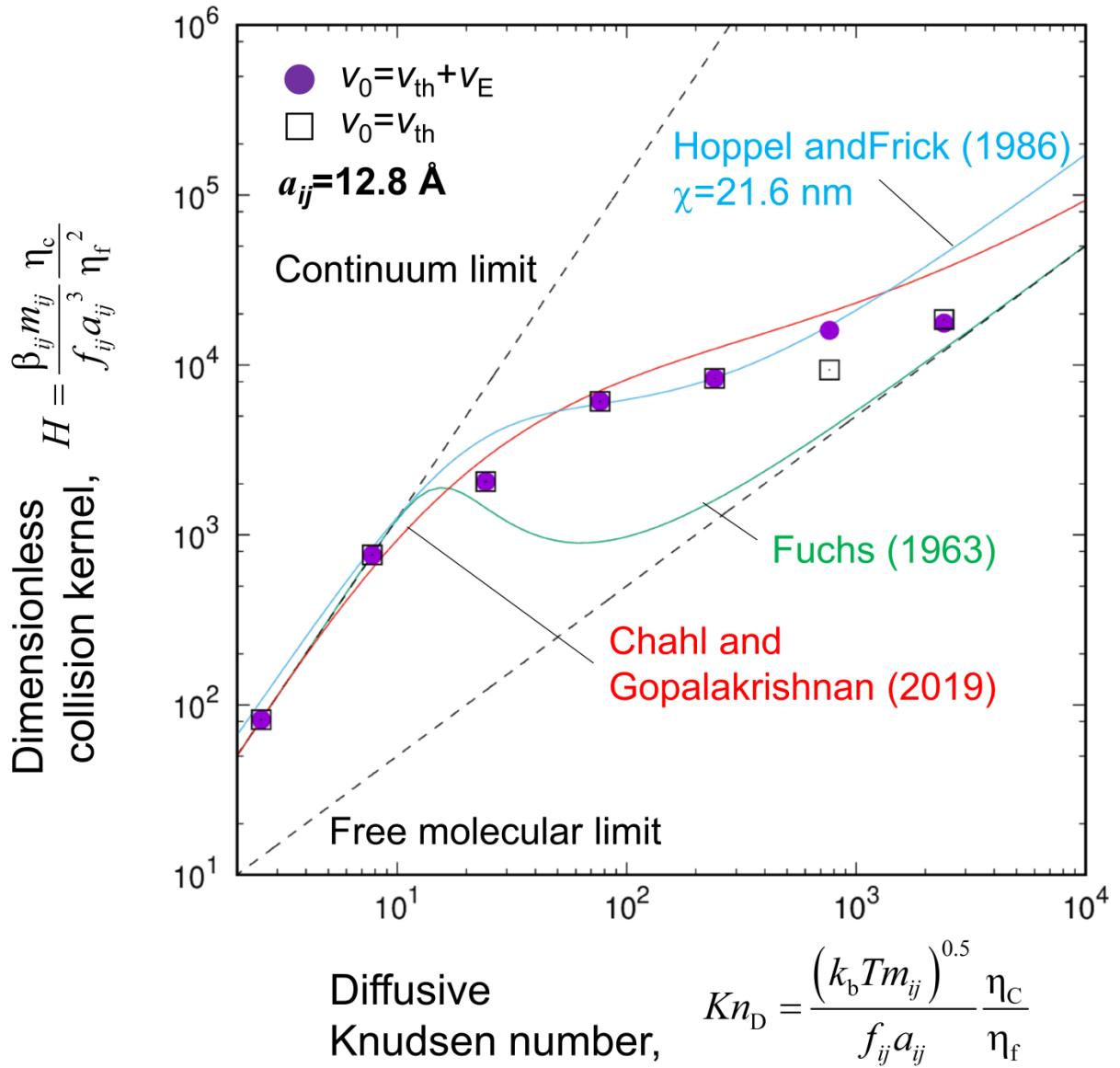

**Figure 9.** The dimensionless recombination rate coefficient as a function of the diffusion Knudsen number. All dimensionless ratios are defined using a collision distance of 12.8 Å. Predictions based on Fuchs[36], and Chahl & Gopalakrishnan[56] (equation 25), and fitting using Hoppel & Frick[38] are also plotted.

## IV. Conclusions

We develop a continuum-molecular dynamics hybrid calculation approach for the ion-ion recombination rate coefficient, which is based upon the limiting sphere approach of Fuchs[36] and Fillipov (1993). Recombination rate coefficient calculations explicitly modeling ion-neutral gas atom (or molecule) collisions are made possible by including two distinct cubic simulation domains around ions whose dimensions are determined by the system temperature and pressure, while the number of gas atoms within each domain is set at $10^3$. Neglect of gas atom-gas atom interactions, variable time stepping, and ensuring that gas atoms only interact with their assigned ion all increase computational speed, which is necessary to simulate the required number of trajectories to determine collision probabilities and recombination rate coefficients. Including all degrees of freedom in ions and inclusion of the neutral gas enables accurate modeling of the outcomes of ion-neutral gas close encounters. We find that ion-neutral collisions fundamentally change the nature of the recombination process; recombining ions can approach one another at grazing angles, in comparison to traditional theories which largely predict "head-on" collisions. For this reason, we advocate adoption and implementation of continuum-molecular dynamics hybrid approaches in lieu of traditional limiting sphere theories where ion motion within the limiting sphere is imprecisely modeled.

Continuum-molecular dynamics calculation predictions, without any fitting, are found to yield recombination rate coefficients in excellent agreement with prior measurements in He gas near atmospheric pressure.[25] Though the test case examined was limited to Helium, by modeling $N_2$ and $O_2$, the approach is also readily extendable to ion-ion recombination in atmospherically relevant conditions and for ion compositions relevant to aerosol charging.[61] Such extensions will require inclusion of induced-dipole potentials between these diatomic gas molecules and ions. Condensable vapor molecules, i.e. water molecules, can also be added to calculations to examine their influence both on ion structure[62] and on the recombination rate coefficient. Furthermore, continuum-molecular dynamics approaches are extendable to examine clustering reactions, condensation, and coagulation; for these processes they can be used for accurate rate calculations at variable temperature and pressure, provided the continuum transport properties of colliding species are known, and provided suitable potentials are developed (either all-atom or coarse-grained) to represent colliding species. As a finally note, beyond the collision rate coefficient, of interest in collisions in clusters is the structure and thermodynamic properties of the collision product, which can be out of thermal equilibrium with its surroundings[63] and for this reason may be anomalously reactive until cooled by subsequent collisions with the background neutral gas. As Coulombically driven collisions occur at substantially elevated speeds, future implementations of continuum-molecular dynamics simulations may afford the opportunity to examine unique Coulombically facilitated chemical reactions for ions, clusters, and charged particles.

**Supporting Information.** Table S1 displays the Pressure, gas simulation conditions, resulting probability, recombination rate coefficient, diffusive Knudsen number, and dimensionless rate coefficient from all simulations performed. Information on the gas atom time steps in simulations are also provided, as are video depictions of ion-ion relative trajectories colliding, not colliding, and orbiting, in the absence of neutral gas.

**Acknowledgements.** C.J.H. acknowledges support from Department of Energy Award No. DE-SC0018202. Computations were performed using computing resources at the Research Center for Computational Science, Okazaki, Japan as well as the Minnesota Supercomputing Institute (MSI) at the University of Minnesota. T. T. was supported by the Hosokawa Powder Technology Foundation. H. H. and T. S. were supported by JST CREST Grant Number JPMJCR18H4, Japan.